\documentclass[prd,11pt,notitlepage,showpacs,showkeys]{revtex4-1}

\usepackage{amsmath,amssymb,bm}
\usepackage{graphicx}
\usepackage{hyperref}
\usepackage{xcolor}

\definecolor{palatinateblue}{rgb}{0.15, 0.23, 0.89}
\definecolor{brightpink}{rgb}{1.0, 0.0, 0.5}
\definecolor{amaranth}{rgb}{0.9, 0.17, 0.31}
\hypersetup{
  colorlinks=true,
  linkcolor=palatinateblue,
  citecolor=brightpink,
  urlcolor=amaranth
}

\newcommand{\vp}{\bm{p}}
\newcommand{\vn}{\bm{\nabla}}

\newcommand{\ii}{\mathrm{i}}

\begin{document}

\title{Three-Dimensional Modified Klein--Gordon Oscillator in Standard and Generalized Doubly Special Relativity}

\author{Abdelmalek Boumali}
\email{abdelmalek.boumali@univ-tebessa.dz}
\email{boumali.abdelmalek@gmail.com}
\affiliation{Laboratory of Theoretical and Applied Physics, Echahid Cheikh Larbi Tebessi University, Algeria}

\author{Nosratollah Jafari}
\email{nosrat.jafari@fai.kz}
\affiliation{Fesenkov Astrophysical Institute, 050020, Almaty, Kazakhstan}
\affiliation{Al-Farabi Kazakh National University, 050040 Almaty, Kazakhstan}
\affiliation{Center for Theoretical Physics, Khazar University, Baku, Azerbaijan}

\date{\today}

\begin{abstract}
Doubly Special Relativity (DSR) augments special relativity by introducing, alongside the invariant speed of light $c$,
a second observer-independent scale typically associated with the Planck regime. At the level of effective wave
equations this principle manifests itself through deformed dispersion relations and energy-dependent spatial operators.
Here we quantify such effects in a prototypical exactly solvable bound-state problem: the three-dimensional
Klein--Gordon oscillator generated by a non-minimal momentum coupling that yields isotropic harmonic confinement while
preserving rotational symmetry. We analyze two standard DSR realizations (Amelino--Camelia and Magueijo--Smolin,
parametrized by an invariant energy scale $k$) as well as a generalized DSR framework based on a first-order expansion
in the Planck length $l_p$. After stationary reduction and separation in spherical coordinates, the eigenfunctions
retain the generalized-Laguerre and spherical-harmonic structure of the undeformed oscillator, whereas DSR deforms the
algebraic quantization condition that relates the principal oscillator number $N=2n+\ell\in\mathbb{N}_0$ to the
relativistic energy. Closed-form spectra are obtained for the standard DSR cases, and perturbative Planck-suppressed
shifts are derived for the generalized model. In all realizations the deformation induces branch-dependent shifts of
both positive- and negative-energy solutions, which increase with excitation and vanish smoothly in the limits
$k\to\infty$ or $l_p\to0$. The main goal of this paper is to extract analytic spectra and Planck-suppressed shifts that enable a direct comparison between different DSR prescriptions in a fully three-dimensional setting.
\end{abstract}


\maketitle

\section{Introduction}
DSR (Doubly Special Relativity) theories extend Einstein’s special relativity by postulating a second observer-independent scale alongside the invariant speed of light $c$: the Planck energy $E_p=\sqrt{\hbar c^5/G}\approx 10^{19},\text{GeV}$. Whereas standard special relativity enforces only the invariance of $c$ between inertial frames, DSR alters the relativistic energy-momentum structure to encode quantum-gravitational effects expected to emerge near the Planck scale. Prominent realizations include the Amelino–Camelia proposal \cite{Amelino-Camelia:2000stu,Amelino-Camelia:2002uql} and the Magueijo–Smolin (MS) construction \cite{Magueijo:2001cr}, which implement distinct deformations while maintaining observer independence of both $c$ and $E_p$ \cite{JafariPhysRevD}.

Although both frameworks introduce the same additional invariant scale $E_p$, they differ markedly in formulation and phenomenology. The Amelino–Camelia (AC) approach is typically derived from deformations of the Lorentz algebra—often via $\kappa$-Poincaré symmetries—and is frequently associated with non-commutative geometric structures. It primarily deforms the momentum sector, yielding modified dispersion relations and, in some realizations, an energy-dependent photon propagation speed. This opens the door to observable signatures such as energy-dependent time-of-flight delays in high-energy photons arriving from distant astrophysical sources. In contrast, the Magueijo–Smolin (MS) model is built through nonlinear (projective) redefinitions of momentum space designed to keep $c$ and $E_p$ exactly invariant. Operationally, the leading deformation tends to enter through the mass/energy sector (at first order in the Planck length), producing characteristic corrections to rest energies and to kinematic thresholds in high-energy processes.

The presence of these two variants reflects different underlying motivations. AC-type deformations are commonly interpreted as arising from quantum-gravity-inspired curvature or nontrivial geometry in momentum space, while MS-type constructions emphasize preserving the relativity principle together with exact invariance of both $c$ and $E_p$ in a modified kinematic setting. From an experimental standpoint, the two can be probed in complementary ways: precision astrophysical timing is generally more sensitive to AC-like dispersion effects, whereas threshold tests in collider, cosmic-ray, or astroparticle contexts can be more revealing of MS-like mass-sector deformations. Although present constraints are still far from direct Planck-scale reach, these channels offer concrete benchmarks for future observational and experimental efforts \cite{Amelino-Camelia:2000stu,Amelino-Camelia:2002uql,Magueijo:2001cr,GuvendiDSRPhysLettB2024,JafariBoumali2025PLB_DiracOscillatorDSR,Coraddu:2009sb,JafariPhysLettB2020,JafariPhysLettB2024,JafariPhysLettB2025}.

More broadly, DSR provides a useful phenomenological framework to explore how Planck-scale physics might imprint itself on quantum systems \cite{JafariPhysLettB2020,JafariPhysLettB2024,JafariPhysLettB2025,JafariPhysRevD,Coraddu:2009sb}.

A complementary perspective emphasizes that DSR is naturally realized in deformations of relativistic symmetry algebras and in nontrivial momentum-space geometries. The $\kappa$-Poincar'e framework provides explicit realizations of deformed Lorentz symmetry and related noncommutative structures. Modern discussions also highlight the ``relative locality'' viewpoint, where an invariant phase space replaces absolute spacetime locality, and momentum-space geometry encodes observable effects \cite{Amelino-Camelia:2000stu,Amelino-Camelia:2002uql}.

Exactly solvable relativistic bound-state models offer a controlled arena for testing how such deformations propagate to quantum spectra. In non-relativistic quantum mechanics, the harmonic oscillator is one of the canonical exactly solvable systems. Within the relativistic domain, the generalization of an oscillator interaction is not unique, yet several relativistic oscillator models remain exactly solvable.
A paradigmatic example is the Dirac oscillator (DO), introduced via a non-minimal substitution in the Dirac equation and widely used as a relativistic analogue of the harmonic oscillator with spin effects \cite{moshinsky1989}. Bruce and Minning introduced a linear interaction within the Klein--Gordon equation known as the Klein--Gordon oscillator (KGO) \cite{bruce1993,Dvoeglazo1994,SiouaneLow2024, ito1967,moshinsky1989,moreno1989,Benitez1990,Bermudez2007,Bermudez2008,Boumali2013,Boumali2015ZNA,Boumaliejtp2015,BoumaliJafariShukirgaliyevSerdouk2025EPJC_KGO_DSR_Thermal,quesne1990,quesne2005}. They incorporated this interaction through the substitution (in operator form)
\begin{equation}
\mathbf{p} \to \mathbf{p}- im\omega\mathbf{r},
\end{equation}
with $\mathbf{r}$ the position operator and $\omega$ the oscillator frequency. Dvoeglazov later discussed the corresponding Hamiltonian/operator-ordering form of this construction and related clarifications \cite{Dvoeglazo1994}.

In this work, we investigate the Klein–Gordon oscillator (KGO) within the doubly special relativity (DSR) framework—an analysis that, to the best of our knowledge, has not been carried out before. Our main purpose is to assess how two representative realizations of DSR, namely the Amelino–Camelia (AC) and Magueijo–Smolin (MS) models, deform the standard KGO dynamics and spectrum \cite{Amelino-Camelia:2000stu,Amelino-Camelia:2002uql,Magueijo:2001cr}. Specifically, we aim (i) to quantify how the deformation scale reshapes the energy spectrum and affects the separation between positive- and negative-energy branches; (ii) to contrast the spectral signatures induced by quantum-gravity-motivated corrections in the AC and MS scenarios; and (iii) to identify the parameter regimes that ensure a real spectrum, while clarifying how the system tends toward the classical limit as the deformation parameter grows.

The remainder of the paper is structured as follows. Section II reviews the undeformed three-dimensional Klein–Gordon oscillator. Section III derives the DSR-modified Klein–Gordon equation in the Amelino–Camelia framework. Section IV develops the corresponding formulation in the Magueijo–Smolin model and provides a detailed comparison of the two sets of results. Section V concludes with a summary and a discussion of implications for Planck-scale phenomenology.

\section{Three-dimensional Klein--Gordon oscillator (undeformed problem)}
The Klein--Gordon equation for a scalar field $\Phi(\bm{r},t)$ reads \cite{Greiner1998QMSpecialChapters,Greiner2001QMIntro,GreinerMueller1994QMSymmetries,Davydov1976QuantumMechanics,Sakurai1967AdvancedQM,SakuraiNapolitano2020ModernQM}
\begin{equation}
\left(\frac{1}{c^2}\frac{\partial^2}{\partial t^2}-\nabla^2+\frac{m^2c^2}{\hbar^2}\right)\Phi(\bm{r},t)=0.
\label{eq:KG_standard}
\end{equation}
The KG oscillator is introduced through the non-minimal substitution
\begin{equation}
\vp \;\to\; \vp - \ii m\omega \bm{r},
\qquad \vp=-\ii\hbar \vn,
\label{eq:nonminimal}
\end{equation}
which generates isotropic harmonic confinement \cite{bruce1993,Dvoeglazo1994}. In three spatial dimensions one uses the operator identity
\begin{equation}
(\vp+\ii m\omega \bm{r})\!\cdot\!(\vp-\ii m\omega \bm{r})
=\vp^{2}+m^2\omega^2 r^2-3m\omega\hbar,
\label{eq:osc_identity}
\end{equation}
with $r=|\bm{r}|$ and $\vn\cdot\bm{r}=3$.

Using the stationary ansatz
\begin{equation}
\Phi(\bm{r},t)=e^{-\ii Et/\hbar}\psi(\bm{r}),
\label{eq:stationary}
\end{equation}
and separation in spherical coordinates,
$\psi(\bm{r})=R_{n\ell}(r)Y_{\ell m}(\theta,\phi)$, one obtains the standard polynomial termination condition
\begin{equation}
N=2n+\ell,\qquad N\in\mathbb{N}_0,\qquad n=0,1,2,\dots,
\label{eq:Ndef}
\end{equation}
so that $N$ is an integer ($N=0,1,2,\dots$). The radial eigenfunctions are proportional to
$r^\ell e^{-\rho/2}L_{n}^{\ell+1/2}(\rho)$ with $\rho=\tfrac{m\omega}{\hbar}r^2$ \cite{Andrews1999}.
A convenient consequence of \eqref{eq:osc_identity} is that the operator
$\vp^{2}+m^2\omega^2 r^2-3m\omega\hbar$ has eigenvalues $2m\omega\hbar\,N$, hence the undeformed spectrum satisfies
\begin{equation}
E_{N}^{(0)\,2}=m^2c^4+2mc^2\hbar\omega\,N,
\qquad
E_N^{(0)}=\pm\sqrt{m^2c^4+2mc^2\hbar\omega\,N}.
\label{eq:E0}
\end{equation}
Equation~(7) shows that the Klein--Gordon oscillator exhibits a two-branch relativistic spectrum
\begin{equation}
E^{(0)}_{N}= \pm\sqrt{m^{2}c^{4}+2mc^{2}\hbar\omega\,N},
\qquad N=2n+\ell\in\mathbb{N}_{0},
\end{equation}
corresponding to the positive-energy (particle) and negative-energy (antiparticle) solutions.
The ground level is $E^{(0)}_{0}=\pm mc^{2}$, and $|E^{(0)}_{N}|$ increases monotonically with $N$.
In the weakly relativistic regime $\hbar\omega N\ll mc^{2}$, the positive branch admits the expansion
\begin{equation}
E^{(0)}_{N}\simeq mc^{2}+\hbar\omega N-\frac{(\hbar\omega N)^{2}}{2mc^{2}}+\cdots,
\end{equation}
so that the nonrelativistic oscillator spacing $\Delta E\simeq \hbar\omega$ emerges above the rest
energy, while the negative branch behaves symmetrically,
$E^{(0)}_{N}\simeq -mc^{2}-\hbar\omega N+\cdots$.

Importantly, since $E^{(0)}_{N}$ depends only on $N=2n+\ell$ (and not separately on $\ell$ or $m$),
each level is degenerate with respect to the angular quantum numbers: for fixed $N$, the allowed
orbital values are $\ell=N,\,N-2,\,\dots\ge 0$, and each $\ell$ carries the usual $(2\ell+1)$ magnetic
degeneracy. The total degeneracy at fixed $N$ is therefore
\begin{equation}
g_{N}=\sum_{\ell=N,N-2,\dots}(2\ell+1)=\frac{(N+1)(N+2)}{2},
\end{equation}
reflecting the isotropy (rotational symmetry) of the three-dimensional oscillator.

\section{DSR-deformed KG oscillator}
In DSR-inspired effective wave equations, deformation effects enter through energy-dependent coefficients multiplying
spatial operators and/or higher-derivative terms. For stationary states these become algebraic functions of $E$, and
the KG-oscillator operator \eqref{eq:osc_identity} remains the central spatial building block. Consequently, the
Laguerre--spherical-harmonic eigenbasis is preserved at leading order, while the spectral relation between $N$ and
$E$ is deformed.

\subsection{Standard DSR realizations}
\subsubsection{Amelino--Camelia type}
A commonly used effective AC-DSR wave equation may be written in the form (see, e.g., Refs.~\cite{Amelino-Camelia:2000stu,Amelino-Camelia:2002uql})
\begin{equation}
\left[
-\frac{1}{c^2}\frac{\partial^2}{\partial t^2}
+\frac{\ii}{k}\frac{\partial}{\partial t}\nabla^2
-(m^2c^2-\nabla^2)
\right]\Phi(\bm{r},t)=0,
\label{eq:AC_mKG}
\end{equation}
where $k$ is an invariant energy scale. After $\Phi=e^{-\ii Et/\hbar}\psi$, the Laplacian term acquires a factor
$\left(1+\tfrac{E}{\hbar k}\right)$. Implementing the oscillator coupling and using the eigenvalue
$2m\omega\hbar\,N$ yields the spectral condition
\begin{equation}
\left(1+\frac{E}{\hbar k}\right)\left(2m\omega\hbar\,N\right)=\frac{E^2}{c^2}-m^2c^2.
\label{eq:AC_condition}
\end{equation}
Solving for $E$ gives two branches, $E_N^{(+)}$ and $E_N^{(-)}$, continuously connected to the undeformed branches:
\begin{equation}
E_N^{\text{(AC)}}=
\frac{m\omega c^2}{k}\,N
\pm
\sqrt{m^2c^4+2m c^2 \hbar\omega\,N
+\left(\frac{m\omega c^2}{k}N\right)^2 }.
\label{eq:AC_spectrum}
\end{equation}

Equation~(\ref{eq:AC_spectrum}) defines a deformed two-branch relativistic spectrum,
\begin{equation}
E^{(\mathrm{AC})}_{N}
=\frac{m\omega c^{2}}{k}\,N
\pm
\sqrt{m^{2}c^{4}+2mc^{2}\hbar\omega\,N+\left(\frac{m\omega c^{2}}{k}\,N\right)^{2}},
\qquad N=2n+\ell\in\mathbb{N}_{0},
\end{equation}
which reduces continuously to the undeformed result (7) in the limit $k\to\infty$.
For large invariant scale $k$ one may expand
\begin{equation}
E^{(\mathrm{AC})}_{N}=\pm E^{(0)}_{N}+\frac{m\omega c^{2}}{k}\,N+\mathcal{O}(k^{-2}),
\end{equation}
so the leading deformation is linear in $N$ and shifts both branches upward by the same absolute
amount (the positive branch increases, whereas the negative branch becomes less negative).
Since $E^{(\mathrm{AC})}_{N}$ depends only on the principal number $N=2n+\ell$, the isotropic
oscillator degeneracy at fixed $N$ is preserved: for a given $N$, $\ell=N,N-2,\dots\ge 0$ and each
$\ell$ carries a $(2\ell+1)$ magnetic degeneracy. Hence the total degeneracy remains
\begin{equation}
g_{N}=\sum_{\ell=N,N-2,\dots}(2\ell+1)=\frac{(N+1)(N+2)}{2}.
\end{equation}

\subsubsection{Magueijo--Smolin type}
An MS-DSR-inspired effective equation may be written as \cite{Magueijo:2001cr,MagueijoSmolin2002_PRL}
\begin{equation}
\left( -\frac{1}{c^2}\frac{\partial^2}{\partial t^2} + \nabla^2 \right)\Phi
=
m^2c^2\left(1- \frac{\ii}{k}\frac{\partial}{\partial t} \right)^2\Phi.
\label{eq:MS_mKG}
\end{equation}
After stationary reduction and oscillator coupling, one obtains a quadratic equation for $E$:
\begin{equation}
\left(\frac{m^2c^4}{k^2}-1\right)E^2
+\frac{2m^2c^4}{k}\,E
+\left(m^2c^4+2m c^2\hbar\omega\,N\right)=0,
\label{eq:MS_quadratic}
\end{equation}
whose two solutions define the positive- and negative-energy branches. The physical branch assignment is fixed by
continuity with Eq.~\eqref{eq:E0} in the limit $k\to\infty$.

Solving Eq.~\eqref{eq:MS_quadratic} yields the closed-form spectrum
\begin{equation}
E_{N}^{\text{(MS)}}=
\frac{\displaystyle \frac{m^2c^4}{k}\pm
\sqrt{\,m^2c^4+2mc^2\hbar\omega\,N\left(1-\frac{m^2c^4}{k^2}\right)}}
{\displaystyle 1-\frac{m^2c^4}{k^2}},
\label{eq:MS_spectrum}
\end{equation}
where the ``$+$'' (``$-$'') sign continuously connects to the positive- (negative-) energy branch
$E_N^{(0)}$ in the undeformed limit $k\to\infty$. For physically motivated scales one typically has
$k\gg mc^2$, so that the denominator in Eq.~\eqref{eq:MS_spectrum} remains close to unity and the deformation is
perturbative.

The MS realization yields the closed-form spectrum in Eq.~(\ref{eq:MS_spectrum}),
\begin{equation}
E^{(\mathrm{MS})}_{N}
=\frac{m^{2}c^{4}}{k}
\pm
\frac{\sqrt{\,m^{2}c^{4}+2mc^{2}\hbar\omega\,N\left(1-\frac{m^{2}c^{4}}{k^{2}}\right)\,}}
{\,1-\frac{m^{2}c^{4}}{k^{2}}\,},
\qquad N=2n+\ell\in\mathbb{N}_{0},
\end{equation}
with the branch assignment fixed by continuity with the undeformed limit $k\to\infty$.
In the physically relevant regime $k\gg mc^{2}$ (so that $1-m^{2}c^{4}/k^{2}\approx 1$),
the leading correction is an $N$-independent offset,
\begin{equation}
E^{(\mathrm{MS})}_{N}=\pm E^{(0)}_{N}+\frac{m^{2}c^{4}}{k}+\mathcal{O}(k^{-2}),
\end{equation}
while the first $N$-dependent deformation enters only at $\mathcal{O}(k^{-2})$ through the factor
$1-m^{2}c^{4}/k^{2}$ under the square root and in the denominator.
As in the undeformed isotropic oscillator, $E^{(\mathrm{MS})}_{N}$ depends only on $N$, so the
degeneracy at fixed $N$ is not lifted:
\begin{equation}
g_{N}=\frac{(N+1)(N+2)}{2}.
\end{equation}

\subsection{Generalized DSR: first-order Planck-length expansion}

Generalized DSR frameworks often encode Planck-scale effects through a deformed dispersion
relation expanded in the Planck length $l_p$.
To first order in $l_p$, a generic modified dispersion relation (MDR) may be written as
\begin{equation}
p_0^{\,2}-p^{2}-2l_p\alpha_2\,p_0^{\,3}+2l_p(\alpha_3-\alpha_1)\,p_0\,p^{2}=m^{2},
\label{eq:gendsr}
\end{equation}
where $\alpha_1,\alpha_2,\alpha_3$ are dimensionless deformation coefficients.
The $\alpha_2$ term represents a purely energy-sector (``time-like'') cubic correction,
while the $(\alpha_3-\alpha_1)$ contribution mixes energy with the spatial momentum sector.
At this order the deformation is Planck suppressed and can be treated perturbatively, so that
standard special relativity is recovered smoothly as $l_p\to 0$.

\subsubsection{1. Operator realization and stationary reduction}

We map $p_0\to i\hbar\,\partial_t/c$ and $p\to -i\hbar\nabla$.
For bound states we employ the stationary ansatz~(5), which reduces time derivatives to powers of
$E$. After implementing the oscillator coupling~(3), the spatial operator $p^2$ is replaced by the
isotropic KG-oscillator operator
\begin{equation}
O\equiv (p+im\omega r)\!\cdot\!(p-im\omega r)=p^{2}+m^{2}\omega^{2}r^{2}-3m\omega\hbar.
\label{eq:operator}
\end{equation}
The operator $O$ is Hermitian and diagonal in the usual three-dimensional oscillator basis, so the
Laguerre--spherical-harmonic eigenfunctions remain unchanged at the level of the present MDR
treatment; the deformation manifests itself through an altered algebraic relation between $N$ and $E$.
The eigenstates are labeled by $N=2n+\ell\in\mathbb{N}_0$, and one has the eigenvalue relation
\begin{equation}
O\,\psi_{n\ell m}=2m\omega\hbar\,N\,\psi_{n\ell m}.
\end{equation}
Because the deformation depends on $O$ only through its eigenvalue, the characteristic isotropic
degeneracy at fixed $N$ (with respect to $\ell$ and $m$) is preserved in this construction.

\subsubsection{2. Energy quantization from the MDR}

With $p_0\to E/c$ and $p^2\to O$, projecting Eq.~(14) onto an eigenstate of $O$ yields an algebraic
equation for $E$. Defining
\begin{equation}
\lambda_N\equiv 2m\omega\hbar\,N,
\tag{17}
\end{equation}
one obtains, to first order in $l_p$,
\begin{equation}
\frac{E^{2}}{c^{2}}-\lambda_N-2l_p\alpha_2\,\frac{E^{3}}{c^{3}}
+2l_p(\alpha_3-\alpha_1)\,\frac{E}{c}\,\lambda_N
= m^{2}c^{2}.
\tag{18}
\end{equation}
Setting $l_p=0$ reproduces the undeformed relation~(7).  It is useful to view Eq.~(18) as
$F(E)=0$ with $F(E)=F_0(E)+l_p F_1(E)$, where
$F_0(E)=E^2/c^2-\lambda_N-m^2c^2$ and
$F_1(E)=-2\alpha_2 E^3/c^3+2(\alpha_3-\alpha_1)(E/c)\lambda_N$.
Writing
\[
E=E^{(0)}_N+l_p\,\delta E,
\]
and expanding to $\mathcal{O}(l_p)$ gives
$F_0(E^{(0)}_N)+l_p\big[F_0'(E^{(0)}_N)\delta E+F_1(E^{(0)}_N)\big]=0$.
Since $F_0(E^{(0)}_N)=0$ and $F_0'(E)=2E/c^2$, one finds
\begin{equation}
\delta E=\alpha_2\,\frac{\big(E^{(0)}_N\big)^2}{c}
-(\alpha_3-\alpha_1)\,c\,\lambda_N,
\tag{19}
\end{equation}
so that
\begin{equation}
E_N \approx E^{(0)}_N+l_p
\left[
\alpha_2\,\frac{\big(E^{(0)}_N\big)^2}{c}
-(\alpha_3-\alpha_1)\,c\,\lambda_N
\right],
\qquad
E^{(0)}_N=\pm\sqrt{m^{2}c^{4}+2mc^{2}\hbar\omega\,N}.
\tag{20}
\end{equation}
Two remarks follow immediately.
(i) The shift $\delta E$ depends on $(E^{(0)}_N)^2$ and on $\lambda_N$, hence it is the \emph{same}
for both branches at fixed $N$; the positive branch is shifted upward by $l_p\delta E$, while the
negative branch becomes less negative by the same absolute amount.
(ii) The correction grows with excitation, through $\lambda_N\propto N$ and through the implicit
$N$-dependence of $(E^{(0)}_N)^2$, so higher oscillator levels are more sensitive to Planck-suppressed
effects within this MDR framework.

\subsubsection{3. MS-type coefficients and first-order resummation}

For the MS-type choice $\alpha_1=0$, $\alpha_2=-1$, $\alpha_3=-1$, Eq.~(18) becomes
\begin{equation}
\frac{E^{2}}{c^{2}}-\lambda_N+2l_p\,\frac{E^{3}}{c^{3}}
-2l_p\,\frac{E}{c}\,\lambda_N
= m^{2}c^{2}.
\tag{21}
\end{equation}
At face value this is cubic in $E$. However, because we are working consistently only to
$\mathcal{O}(l_p)$, one can simplify the cubic term without solving the full cubic equation:
inside the $l_p$-suppressed contribution, we may replace $E^2/c^2$ by its zeroth-order value
$m^2c^2+\lambda_N$ up to $\mathcal{O}(l_p)$ corrections. Concretely,
\[
\frac{E^{3}}{c^{3}}=\frac{E}{c}\left(\frac{E^{2}}{c^{2}}\right)
\simeq \frac{E}{c}\left(m^{2}c^{2}+\lambda_N\right),
\]
which collapses Eq.~(21) to an $\mathcal{O}(l_p)$-accurate quadratic equation:
\begin{equation}
E^{2}+2l_p m^{2}c^{3}E-\Big(m^{2}c^{4}+c^{2}\lambda_N\Big)=0.
\tag{22}
\end{equation}
Therefore,
\begin{equation}
E^{(\mathrm{gen,MS})}_N
= -l_p m^{2}c^{3}\pm \sqrt{\,l_p^{2}m^{4}c^{6}+m^{2}c^{4}+c^{2}\lambda_N\,}
= -l_p m^{2}c^{3}\pm \sqrt{\,l_p^{2}m^{4}c^{6}+m^{2}c^{4}+2mc^{2}\hbar\omega\,N\,}.
\tag{23}
\end{equation}
The $l_p\to 0$ limit reproduces $E^{(0)}_N$ immediately. The MS-type generalized spectrum exhibits
a branch-independent offset $-l_p m^{2}c^{3}$ plus a modified square-root term; at first order in $l_p$
this reproduces the perturbative shift pattern encoded in Eq.~(20) with the MS-type coefficients.

\subsubsection{4. AC-type coefficients and linear shift}

For the AC-type choice $\alpha_1=-\tfrac{1}{2}$, $\alpha_2=0$, $\alpha_3=-1$, one has
$\alpha_3-\alpha_1=-\tfrac{1}{2}$ and the cubic term in Eq.~(18) is absent.
In this case Eq.~(20) yields the compact first-order result
\begin{equation}
E^{(\mathrm{gen,AC})}_N \approx \pm\sqrt{m^{2}c^{4}+2mc^{2}\hbar\omega\,N}
+ l_p\,m\hbar\omega\,c^{3}\,N,
\tag{24}
\end{equation}
i.e.\ a Planck-suppressed correction growing linearly with $N$.
As in the undeformed oscillator, the dependence on $N$ alone implies that the $(\ell,m)$ degeneracy
at fixed $N$ is not lifted within this first-order generalized AC-type implementation.
Moreover, the strictly linear scaling in $N$ provides a clean qualitative marker that parallels the
leading large-$k$ behavior found in the standard AC realization.

\section{Results and discussion}
In this section we discuss the qualitative and quantitative signatures of DSR-induced deformations in the KG-oscillator
spectrum. The two plots summarize the dependence of the energy eigenvalues on the principal oscillator number
$N=2n+\ell\in\mathbb{N}_0$ and isolate the corresponding relative shifts with respect to the undeformed spectrum.

\begin{figure}[t]
\centering
\includegraphics[width=0.85\linewidth]{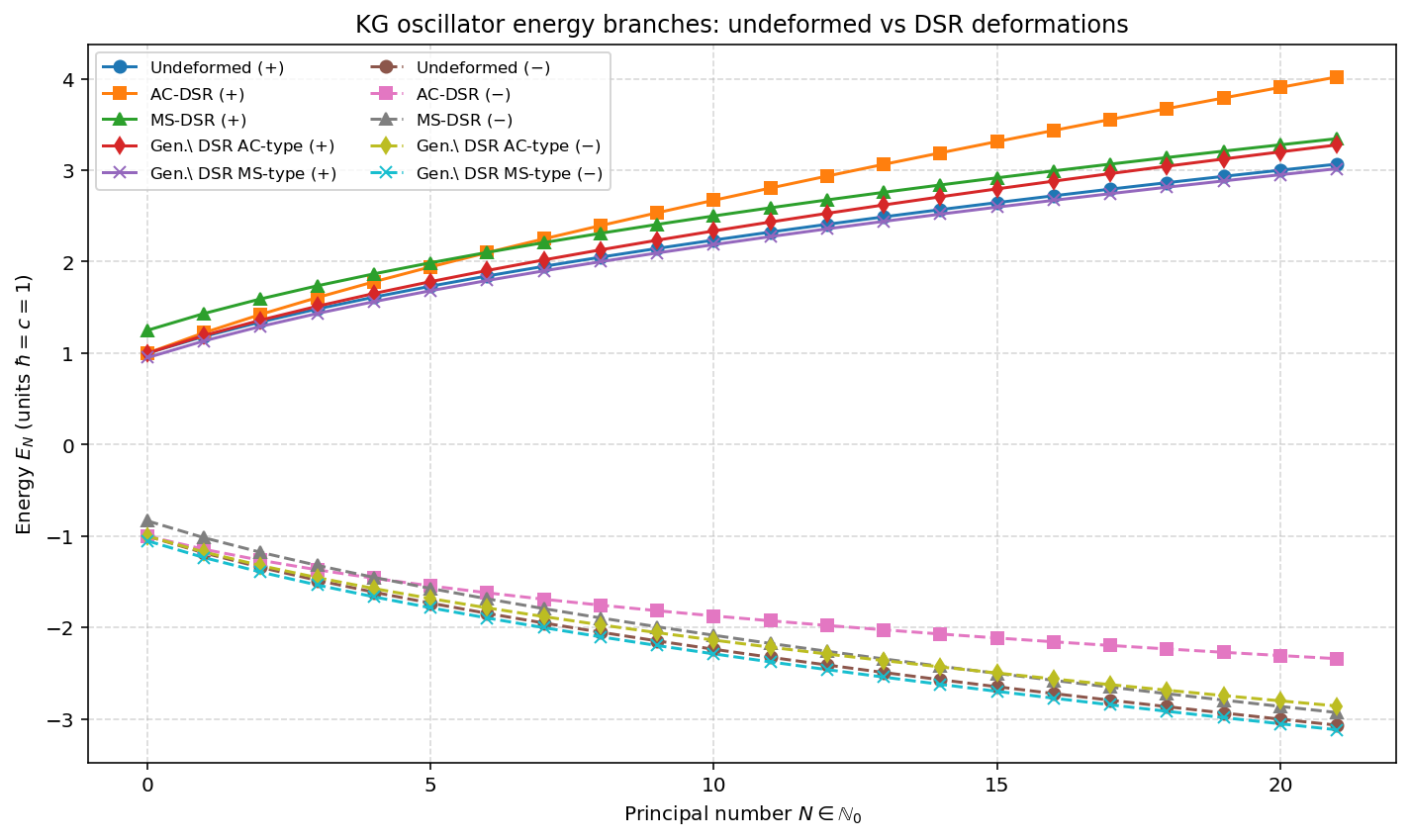}
\caption{
Energy spectrum $E_N$ as a function of the principal oscillator number $N$ for the Klein--Gordon oscillator,
showing both branches: the positive-energy branch $E_N^{(+)}$ (solid lines) and the negative-energy branch
$E_N^{(-)}$ (dashed lines). The undeformed result $E_N^{(0)}=\pm\sqrt{m^2c^4+2mc^2\hbar\omega\,N}$ is compared with
representative DSR-deformed realizations (standard AC-DSR, standard MS-DSR, and generalized DSR in AC/MS-type
first-order $l_p$ expansions). Here $N\in\mathbb{N}_0$ is an integer, $N=0,1,2,\dots$.
}
\label{fig:levels}
\end{figure}

\begin{figure}[t]
\centering
\includegraphics[width=0.85\linewidth]{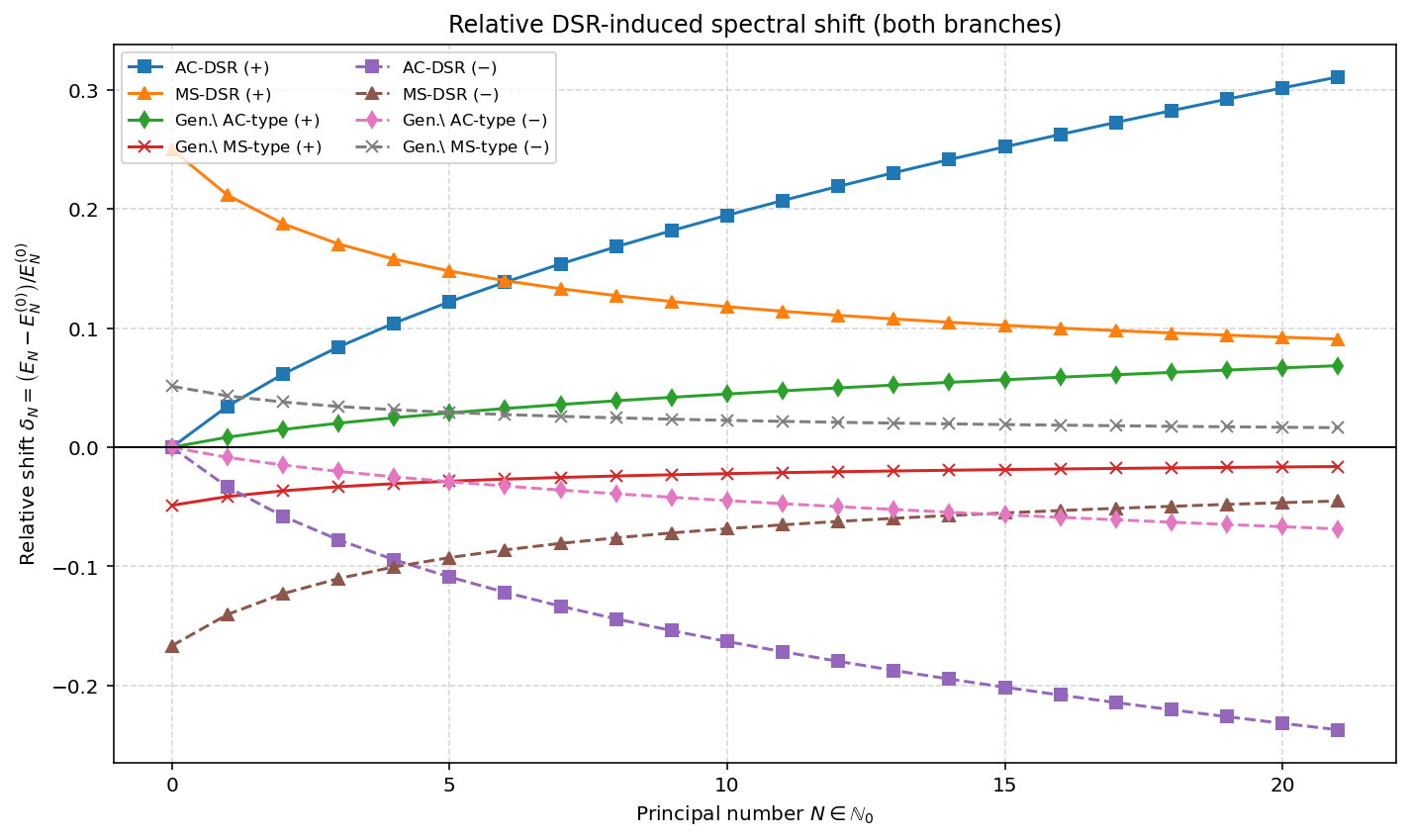}
\caption{
Relative spectral shift $\delta_N\equiv\big(E_N-E_N^{(0)}\big)/E_N^{(0)}$ as a function of $N$ for both branches.
Solid (dashed) curves correspond to the positive- (negative-) energy branch, and the denominator uses the matching
undeformed branch $E_N^{(0)}$ so that $E_N^{(0)}<0$ for the negative branch. The figure isolates the deformation
signal and highlights its growth with excitation number $N\in\mathbb{N}_0$ (integer $N=0,1,2,\dots$).
}
\label{fig:shift}
\end{figure}

\subsection{Robust structural features}
A central technical point is that the isotropic KG-oscillator operator $O$ in Eq.~\eqref{eq:operator} remains
the relevant spatial building block across the deformed models considered in Sec.~III. Since $O$ is diagonal
in the standard oscillator basis, separability in spherical coordinates is preserved and the radial eigenfunctions
retain the generalized-Laguerre form, while the angular dependence remains encoded in spherical harmonics
\cite{Andrews1999}. Consequently, DSR affects the spectrum predominantly through an \emph{algebraic} deformation of the
quantization condition relating $N$ to the energy. In particular, because the deformed energy depends only on $N$,
the characteristic degeneracy of the isotropic oscillator with respect to $(\ell,m)$ at fixed $N$ is not lifted at the
level studied here.

\subsection{Standard DSR: AC versus MS realizations}
The AC-DSR realization modifies the stationary reduction by an energy-dependent factor multiplying the spatial
operator, leading to the condition \eqref{eq:AC_condition} and to the spectrum \eqref{eq:AC_spectrum}. For large
invariant scale $k$ one finds the expansion
\begin{equation}
E_N^{\text{(AC)}}=\pm E_N^{(0)}+\frac{m\omega c^2}{k}\,N+O(k^{-2}),
\label{eq:AC_expansion}
\end{equation}
i.e.\ the leading correction is linear in the excitation label $N$ and shifts \emph{both} branches upward by the same
absolute amount. As a result, the positive-energy branch is increased, whereas the negative-energy branch becomes less
negative. This asymmetry is clearly reflected in Fig.~\ref{fig:shift}: because the relative shift is normalized by
$E_N^{(0)}$, the sign of $\delta_N$ differs between the two branches even though the absolute shift at $O(k^{-1})$ is
the same.

By contrast, in the MS-DSR realization the explicit spectrum \eqref{eq:MS_spectrum} shows that the leading correction
at large $k$ is dominated by an $N$-independent offset,
\begin{equation}
E_N^{\text{(MS)}}=\pm E_N^{(0)}+\frac{m^2c^4}{k}+O(k^{-2}),
\label{eq:MS_expansion}
\end{equation}
while the first $N$-dependent deformation appears only at $O(k^{-2})$ through the factor
$\bigl(1-m^2c^4/k^2\bigr)$ under the square root in Eq.~\eqref{eq:MS_spectrum}. Thus, AC- and MS-type realizations can
be distinguished not only by the magnitude of the deformation but also by the scaling of the leading correction with
the excitation number: AC-type deformations generate an $O(N/k)$ correction, whereas MS-type deformations yield an
$O(1/k)$ offset plus subleading $N$-dependent terms.

\subsection{Generalized DSR: coefficient dependence and scaling}
In the generalized framework based on the modified dispersion relation \eqref{eq:gendsr}, the first-order
energy shift in Eq.~\eqref{eq:operator} makes the dependence on both the excitation label and deformation
coefficients explicit. Two qualitatively different contributions can be identified: (i) a term controlled by $\alpha_2$
that scales with $\left(E_N^{(0)}\right)^2$ and hence grows with $N$ through the undeformed relativistic dispersion,
and (ii) a term proportional to $(\alpha_3-\alpha_1)\lambda_N$ that grows linearly with $N$ through
$\lambda_N=2m\omega\hbar N$. Depending on the signs and relative sizes of these coefficients, the generalized model can
either enhance or partially cancel the net deformation-induced shift. The MS-type and AC-type coefficient choices
in Sec.~III illustrate two limiting patterns: a largely branch-independent offset in Eq.~\eqref{eq:MS_spectrum} and
a strictly linear-in-$N$ Planck-suppressed correction in Eq.~\eqref{eq:AC_spectrum}.

\subsection*{The novelty in generalized DSR compared to standard DSR?}

The standard DSR realizations studied here (AC and MS) are \emph{model-fixed} prescriptions:
once an effective wave equation is chosen, the deformation is essentially controlled by a single
invariant energy scale $k$, and one obtains a characteristic leading behavior (linear-in-$N$ for AC,
and mainly an $N$-independent offset for MS at large $k$)
\cite{Amelino-Camelia:2000stu,Magueijo:2001cr,KowalskiGlikman2005IntroDSR,FreidelKowalskiGlikmanSmolin2011_PRD}.
This is precisely why AC and MS can be discriminated by the scaling of their leading corrections
with the excitation label $N$.

By contrast, the generalized DSR framework introduced in Sec.~III\,B is \emph{model-flexible} and
\emph{coefficient-resolved}. Instead of committing to one specific nonlinear realization, it starts
from a first-order Planck-length expansion of a generic modified dispersion relation (MDR),
\begin{equation}
p_0^2 - p^2 - 2l_p\alpha_2 p_0^3 + 2l_p(\alpha_3-\alpha_1)p_0 p^2 = m^2 ,
\end{equation}
which cleanly separates an energy-sector cubic correction (controlled by $\alpha_2$) from a mixed
energy--momentum correction (controlled by $\alpha_3-\alpha_1$). At this order, the deformation
is Planck suppressed and SR is recovered smoothly as $l_p\to 0$.
This kind of systematic MDR expansion is standard in effective descriptions of Planck-scale
kinematics and their geometric interpretations \cite{GirelliLiberatiSindoni2007_FinslerMDR}.

The novelty is therefore \emph{not} merely quantitative (another small correction), but structural:
the generalized approach provides a \emph{unified parameterization} that (i) interpolates between
distinct deformation patterns through $(\alpha_1,\alpha_2,\alpha_3)$, (ii) makes the \emph{scaling
origin} of spectral shifts explicit by isolating contributions that scale like $(E_N^{(0)})^2$
and like $\lambda_N\propto N$, and (iii) allows for \emph{enhancement or partial cancellation}
between these contributions depending on the coefficient signs and magnitudes. This coefficient
control is absent in a single fixed standard-DSR wave equation.
Related “beyond-a-single-model” kinematics, where the relativity principle constrains sets of
coefficients across dispersion and other kinematic structures, has been developed in a systematic
way in the deformed-kinematics literature \cite{Mercati2012_RKBeyondSR}.

Operationally, the generalized MDR viewpoint also fits naturally within the momentum-space
geometry perspective: different coefficient choices can be understood as encoding different
features (e.g., curvature and related structures) in momentum space, which is the conceptual basis
behind relative-locality formulations \cite{FreidelKowalskiGlikmanSmolin2011_PRD,AmelinoCameliaEtAl2011_RelLocality}.
In the present isotropic oscillator problem, this also clarifies why the spatial eigenbasis remains
unchanged at the order considered here: the deformation enters the bound-state problem only through
the eigenvalue of the isotropic KG-oscillator operator, so separability and the $(\ell,m)$ degeneracy
at fixed $N$ persist, while DSR acts mainly as an algebraic reassignment of energies to the same
spatial states. Finally, by expanding in $l_p$, one obtains a systematic route to higher-order
corrections, which can be used to test when genuinely new effects may arise beyond first order.

\subsection{Physical interpretation and limitations}
From an effective-wave-equation viewpoint, the deformed stationary reductions may be interpreted as inducing an
energy-dependent renormalization of the kinetic operator and, consequently, an energy-dependent mapping between the
oscillator quantum number $N$ and the relativistic invariant $E^2-m^2c^4$. Because the eigenfunctions are unchanged at
the level treated here, the deformation can be viewed as a reassignment of energies to otherwise identical spatial
states. This feature makes the KG oscillator a transparent benchmark for isolating kinematical effects of DSR-type
models.

Two limitations are worth emphasizing. First, the standard DSR wave equations used here are phenomenological
realizations designed to capture the leading imprint of deformed dispersion relations; different operator orderings or
higher-order terms may modify quantitative details without necessarily changing the qualitative branch structure.
Second, the generalized analysis is restricted to first order in $l_p$; higher-order corrections may introduce
additional nonlinearities in $E$ and could, in more elaborate models, couple to angular momentum in a way that lifts
degeneracies. Extensions in these directions, as well as the inclusion of external fields or curved backgrounds, offer
natural avenues for future work.

\section{V. CONCLUSION}

We analyzed the three-dimensional Klein--Gordon oscillator under Planck-scale deformations in
two complementary ways: (i) standard DSR realizations of the Amelino--Camelia and
Magueijo--Smolin type, characterized by an invariant energy scale $k$
\cite{Amelino-Camelia:2000stu,Magueijo:2001cr},
and (ii) a generalized DSR framework built from a first-order expansion in the Planck length $l_p$
of a generic modified dispersion relation (MDR), consistent with common effective-kinematics
approaches \cite{GirelliLiberatiSindoni2007_FinslerMDR}. In all cases the oscillator was introduced
via the same non-minimal momentum coupling, and the stationary reduction plus spherical separation
allowed us to derive closed or perturbative spectral relations between the principal oscillator
number $N=2n+\ell$ and the relativistic energy.

A robust outcome is that, for the isotropic setting treated here, DSR does not obstruct
separability: the Laguerre--spherical-harmonic eigenfunctions remain intact at leading order, and
DSR manifests mainly through an algebraic deformation of the $N \leftrightarrow E$ quantization
condition. Consequently, the characteristic degeneracy $(\ell,m)$ at fixed $N$ is preserved within
the present approximations, while both positive and negative-energy branches experience
deformation-induced changes that vanish smoothly as $k\to\infty$ or $l_p\to 0$.

Beyond reproducing these generic features, our results provide a clear discriminator between
standard DSR prescriptions: AC-type deformations generate a leading correction that scales
linearly with the excitation label, $\mathcal{O}(N/k)$, whereas MS-type deformations yield a
dominant $N$-independent offset at $\mathcal{O}(1/k)$ with $N$-dependence entering at higher
order \cite{KowalskiGlikman2005IntroDSR}.
The generalized MDR approach adds a further level of novelty by introducing
coefficient-resolved control: at first order in $l_p$, the energy shift separates into distinct
contributions governed by $\alpha_2$ (energy-sector) and by $(\alpha_3-\alpha_1)$ (mixed
energy--momentum), making the scaling with $(E_N^{(0)})^2$ and with $\lambda_N\propto N$
explicit and allowing for enhancement or partial cancellation depending on the coefficient choices.
This “coefficient-consistent” viewpoint aligns with systematic developments of deformed kinematics
beyond a single DSR map \cite{Mercati2012_RKBeyondSR}.

In addition, generalized DSR is naturally compatible with the momentum-space geometry/relative
locality interpretation, where different deformations correspond to different geometric data on
momentum space \cite{FreidelKowalskiGlikmanSmolin2011_PRD,AmelinoCameliaEtAl2011_RelLocality}.
In this sense, generalized DSR acts as a unified, model-flexible umbrella that can emulate
AC/MS-like behaviors at leading order while simultaneously enabling broader phenomenological scans
without committing to a single fixed deformation map.

The analytic spectra and relative shifts obtained here therefore constitute a clean benchmark for
probing Planck-suppressed kinematics in relativistic bound-state systems. Natural next steps are
to extend the MDR expansion beyond first order in $l_p$, explore anisotropic or interaction-induced
deformations that could lift degeneracies, and incorporate external fields or curved backgrounds
where DSR effects may compete or combine with geometric contributions.

\begin{acknowledgments}
The Science Committee of the Ministry of Science and Higher Education of the Republic of Kazakhstan funds this research
(Grant No.\ AP19677351 and Program No.\ BR21881880).
\end{acknowledgments}

\bibliographystyle{apsrev4-1}
\bibliography{referencearticle}

\end{document}